\documentclass[journal]{IEEEtran}
\usepackage{url}
\usepackage[utf8]{inputenc}
\usepackage{xcolor}
\usepackage{amsmath}
\usepackage{amssymb}

\usepackage[acronyms,nonumberlist,nopostdot,nomain,nogroupskip]{glossaries}
\usepackage{tablefootnote}
\usepackage{booktabs}
\usepackage{tabularx}
\usepackage{tikz}
\usepackage{pgfplots}
\pgfplotsset{compat=newest} 
\pgfplotsset{plot coordinates/math parser=false} 
\newlength\fheight
\newlength\fwidth
\usetikzlibrary{plotmarks,patterns,decorations.pathreplacing,backgrounds,calc,arrows,arrows.meta,matrix,spy,decorations.fractals,shapes.misc}
\usepgfplotslibrary{patchplots,groupplots}
\usepackage{tikzscale}

\usepackage{multirow}
\usepackage{tkz-kiviat}

\usepackage[font=scriptsize]{subcaption}
\usepackage[font=footnotesize]{caption}

\usepackage{mathtools}

\usepackage{dblfloatfix}    
\usepackage{colortbl}

\usepackage{makecell}
\usepackage{diagbox}
\usepackage{tikz-qtree}
\usetikzlibrary{trees} 

\newacronym{3gpp}{3GPP}{3rd Generation Partnership Project}
\newacronym{adc}{ADC}{Analog to Digital Converter}
\newacronym{5g}{5G}{5th generation}
\newacronym{aimd}{AIMD}{Additive Increase Multiplicative Decrease}
\newacronym{am}{AM}{Acknowledged Mode}
\newacronym{amc}{AMC}{Adaptive Modulation and Coding}
\newacronym{aqm}{AQM}{Active Queue Management}
\newacronym{awgn}{AGWN}{Additive White Gaussian Noise}
\newacronym{balia}{BALIA}{Balanced Link Adaptation}
\newacronym{bdp}{BDP}{Bandwidth-Delay Product}
\newacronym{bf}{BF}{beamforming}
\newacronym{cc}{CC}{Congestion Control}
\newacronym{cdf}{CDF}{Cumulative Distribution Function}
\newacronym{cn}{CN}{Core Network}
\newacronym{cav}{CAV}{Connected and Autonomous Vehicles}
\newacronym{cqi}{CQI}{Channel Quality Information}
\newacronym{cp}{CP}{Control Plane}
\newacronym{csirs}{CSI-RS}{Channel State Information - Reference Signal}
\newacronym{dc}{DC}{Dual Connectivity}
\newacronym{dce}{DCE}{Direct Code Execution}
\newacronym{dci}{DCI}{Downlink Control Information}
\newacronym{dl}{DL}{Downlink}
\newacronym{dmr}{DMR}{Deadline Miss Ratio}
\newacronym{dmrs}{DMRS}{DeModulation Reference Signal}
\newacronym{e2e}{E2E}{End-to-End}
\newacronym{ecn}{ECN}{Explicit Congestion Notification}
\newacronym{edf}{EDF}{Earliest Deadline First}
\newacronym{enb}{eNB}{evolved Node Base}
\newacronym{epc}{EPC}{Evolved Packet Core}
\newacronym{es}{ES}{Edge Server}
\newacronym{fdma}{FDMA}{Frequency Division Multiple Access}
\newacronym{fdd}{FDD}{Frequency Division Duplexing}
\newacronym[firstplural=Radio Access Technologies (RATs)]{rat}{RAT}{Radio Access Technology}
\newacronym[firstplural=Radio  Technologies (RTs)]{rt}{RT}{Radio Technology}
\newacronym{fs}{FS}{Fast Switching}
\newacronym{ftp}{FTP}{File Transfer Protocol}
\newacronym{gnb}{gNB}{Next Generation Node Base}
\newacronym{harq}{HARQ}{Hybrid Automatic Repeat reQuest}
\newacronym{hetnet}{HetNet}{Heterogeneous Network}
\newacronym{hh}{HH}{Hard Handover}
\newacronym{hol}{HOL}{Head-of-Line}
\newacronym{ia}{IA}{Initial Access}
\newacronym{imt}{IMT}{International Mobile Telecommunication}
\newacronym{iot}{IoT}{Internet of Things}
\newacronym{los}{LOS}{Line of Sight}
\newacronym{lte}{LTE}{Long Term Evolution}
\newacronym{m2m}{M2M}{Machine to Machine}
\newacronym{mac}{MAC}{Medium Access Control}
\newacronym{mc}{MC}{Multi-Connectivity}
\newacronym{mcs}{MCS}{Modulation and Coding Scheme}
\newacronym{mec}{MEC}{Mobile Edge Cloud}
\newacronym{mi}{MI}{Mutual Information}
\newacronym{mimo}{MIMO}{Multiple Input, Multiple Output}
\newacronym{mmwave}{mmWave}{millimeter wave}
\newacronym{mptcp}{MPTCP}{Multipath TCP}
\newacronym{mr}{MR}{Maximum Rate}
\newacronym{mss}{MSS}{Maximum Segment Size}
\newacronym{mtd}{MTD}{Machine-Type Device}
\newacronym{mtu}{MTU}{Maximum Transmission Unit}
\newacronym{nfv}{NFV}{Network Function Virtualization}
\newacronym{nlos}{NLOS}{Non Line of Sight}
\newacronym{nr}{NR}{New Radio}
\newacronym{ofdm}{OFDM}{Orthogonal Frequency Division Multiplexing}
\newacronym{pdcch}{PDCCH}{Physical Downlonk Control Channel}
\newacronym{pdcp}{PDCP}{Packet Data Convergence Protocol}
\newacronym{pdsch}{PDSCH}{Physical Downlink Shared Channel}
\newacronym{pdu}{PDU}{Packet Data Unit}
\newacronym{pf}{PF}{Proportional Fair}
\newacronym{pgw}{PGW}{Packet Gateway}
\newacronym{phy}{PHY}{Physical}
\newacronym{pbch}{PBCH}{Physical Broadcast Channel}
\newacronym[plural=\gls{mme}s,firstplural=Mobility Management Entities (MMEs)]{mme}{MME}{Mobility Management Entity}
\newacronym{prb}{PRB}{Physical Resource Block}
\newacronym{pss}{PSS}{Primary Synchronization Signal}
\newacronym{pucch}{PUCCH}{Physical Uplink Control Channel}
\newacronym{pusch}{PUSCH}{Physical Uplink Shared Channel}
\newacronym{rach}{RACH}{Random Access Channel}
\newacronym{ran}{RAN}{Radio Access Network}
\newacronym{red}{RED}{Random Early Detection}
\newacronym{rf}{RF}{Radio Frequency}
\newacronym{rlc}{RLC}{Radio Link Control}
\newacronym{rlf}{RLF}{Radio Link Failure}
\newacronym{rrc}{RRC}{Radio Resource Control}
\newacronym{rrm}{RRM}{Radio Resource Management}
\newacronym{rr}{RR}{Round Robin}
\newacronym{rs}{RS}{Remote Server}
\newacronym{rsrp}{RSRP}{Reference Signal Received Power}
\newacronym{rss}{RSS}{Received Signal Strength}
\newacronym{rtt}{RTT}{Round Trip Time}
\newacronym{rw}{RW}{Receive Window}
\newacronym{rx}{RX}{Receiver}
\newacronym{sa}{SA}{standalone}
\newacronym{sack}{SACK}{Selective Acknowledgment}
\newacronym{sap}{SAP}{Service Access Point}
\newacronym{sch}{SCH}{Secondary Cell Handover}
\newacronym{scoot}{SCOOT}{Split Cycle Offset Optimization Technique}
\newacronym{sdma}{SDMA}{Spatial Division Multiple Access}
\newacronym{sinr}{SINR}{Signal to Interference plus Noise Ratio}
\newacronym{sm}{SM}{Saturation Mode}
\newacronym{snr}{SNR}{Signal to Noise Ratio}
\newacronym{son}{SON}{Self-Organizing Network}
\newacronym{ss}{SS}{Synchronization Signal}
\newacronym{srs}{SRS}{Sounding Reference Signal}
\newacronym{sss}{SSS}{Secondary Synchronization Signal}
\newacronym{tb}{TB}{Transport Block}
\newacronym{tcp}{TCP}{Transmission Control Protocol}
\newacronym{tdd}{TDD}{Time Division Duplexing}
\newacronym{tdma}{TDMA}{Time Division Multiple Access}
\newacronym{tfl}{TfL}{Transport for London}
\newacronym{tm}{TM}{Transparent Mode}
\newacronym{trp}{TRP}{Transmitter Receiver Pair}
\newacronym{tti}{TTI}{Transmission Time Interval}
\newacronym{ttt}{TTT}{Time-to-Trigger}
\newacronym{tx}{TX}{Transmitter}
\newacronym{ue}{UE}{User Equipment}
\newacronym{ul}{UL}{Uplink}
\newacronym{uml}{UML}{Unified Modeling Language}
\newacronym{um}{UM}{Unacknowledged Mode}
\newacronym{utc}{UTC}{Urban Traffic Control}
\newacronym{vm}{VM}{Virtual Machine}
\newacronym{rsrq}{RSRQ}{Reference Signal Received Quality}
\newacronym{rssi}{RSSI}{Received Signal Strength Indicator}
\newacronym{crs}{CRS}{Cell Reference Signal}
\newacronym{v2v}{V2V}{Vehicle-to-Vehicle}
\newacronym{v2i}{V2I}{Vehicle-to-Infrastructure}
\newacronym{v2n}{V2N}{Vehicle-to-Network}
\newacronym{v2x}{V2X}{Vehicle-to-Everything}
\newacronym{vn}{VN}{Vehicular Node}
\newacronym{dsrc}{DSRC}{Dedicated Short Range Communication}
\newacronym{ci}{CI}{context information}
\newacronym{voi}{VoI}{value of information}
\newacronym{gps}{GPS}{Global Positioning System}
\newacronym{qos}{QoS}{Quality of Service}
\newacronym{ml}{ML}{Machine Learning}
\newacronym{ahp}{AHP}{Analytic Hierarchy Process}
\newacronym{lidar}{LIDAR}{Light Detection and Ranging}
\newacronym{sumo}{SUMO}{Simulation of Urban MObility}
\newacronym{rsu}{RSU}{Road Side Unit}
\usepackage{cite}
\usepackage{empheq}
\usepackage{afterpage}

\usepackage{siunitx}
\makeglossaries

\def \eqPLLTE{
\begin{equation}
P_{\rm LOS}(d) = \min \Big( \tfrac{0.018}{d},1 \Big) \left[ 1-\exp\left(\frac{-d}{0.063}\right)\right]+
\exp\left(\frac{-d}{0.063}\right),
\end{equation}
}

\IEEEoverridecommandlockouts
\IEEEpubid{%
\makebox[\columnwidth]{ISBN 978-3-903176-05-8~\copyright~2018 IFIP \hfill}%
\hspace{\columnsep}%
\makebox[\columnwidth]{ }%
}

\begin{document}

\title{Performance Study of LTE and mmWave in Vehicle-to-Network Communications}

\author{\IEEEauthorblockN{Marco Giordani$^{\circ }$, Andrea Zanella$^{\circ }$, Takamasa Higuchi$^{\dagger }$, Onur Altintas$^{\dagger}$, Michele Zorzi$^{\circ }$}
\IEEEauthorblockA{\\
$^{\circ }$Consorzio Futuro in Ricerca (CFR) and  University of Padova, Italy, 
 Email:{\{giordani,zanella,zorzi\}@dei.unipd.it, }\\
$^{\dagger}$TOYOTA InfoTechnology Center, Mountain View, CA, USA,
Email:{\{ta-higuchi,onur\}@us.toyota-itc.com }}}

\maketitle

\begin{abstract}
A key enabler for the emerging autonomous and cooperative driving services is high-throughput and reliable Vehicle-to-Network (V2N) communication. In this respect, the millimeter wave (mmWave) frequencies hold great promises because of the large available bandwidth which may provide the required link capacity. 
However, this potential is hindered by the challenging propagation characteristics of high-frequency channels and the dynamic topology of the  vehicular scenarios, which affect the reliability of the connection.   
Moreover, mmWave transmissions typically leverage beamforming gain to compensate for the increased path loss experienced at high frequencies. This, however, requires fine alignment of the transmitting and  receiving beams, which may be difficult in vehicular scenarios.
Those limitations may undermine the performance of V2N communications and pose new challenges for proper vehicular communication design. In this paper, we study by simulation the practical feasibility of some mmWave-aware strategies to support V2N, in comparison to the traditional LTE  connectivity below 6 GHz.
The results show that the orchestration among different radios represents a viable solution to enable both high-capacity and robust V2N~communications.
\end{abstract}


\section{Introduction} 
\label{sec:introduction}
Emerging advanced automotive services aim at enhancing the road safety, reducing the environmental impact of the traffic and the fuel consumption, and supporting infotainment applications~\cite{3GPP_22186}. Such services can  greatly benefit from \gls{v2v} and \gls{v2n} communications, which would make it possible to realize cooperative perception and maneuvering, e.g., by enabling the sharing of  sensor data and driving decisions among vehicles and infrastructure nodes ~\cite{bookV2X}.
Currently, the principal wireless technology supporting \gls{v2v} communications is the  IEEE 802.11p standard, which offers  data exchange at a nominal rate ranging from 6 to 27 Mbps within a range of a few hundreds of meters~\cite{DSRC}. 
\gls{v2n} communications, instead, presently use~the  \gls{lte} connectivity below 6 GHz, enabling a data rate of up to 100 Mbps in high mobility~scenarios~\cite{araniti2013LTE}.
 
However, the data generation rate of vehicles can soon reach the order of terabytes per driving hour, exceeding the capacity of traditional technologies for vehicular communications \cite{Heath_surveyV2X}.
In this regard, the \gls{3gpp} has recently investigated new \glspl{rt} as enablers for the performance requirements of future automotive services, including the \gls{mmwave} spectrum – roughly above 10 GHz~\cite{magazine2016_Heath}.
Besides the large~bandwidths available at such frequencies, the small size of antennas at mmWaves makes it practical to build very large antenna arrays and obtain high gains by \gls{bf}, thereby guaranteeing extremely high transmission speeds~\cite{rappaport2014millimeter}.
On the other hand,  the severe isotropic path loss and the harsh propagation characteristics of the \gls{mmwave} channel, as well as the dynamic topology of the vehicular networks, have raised many concerns about the applicability of this technology to a vehicular context.
Moreover, \gls{mmwave} links are typically directional and require precise alignment of the transmitter and receiver beams to maintain connectivity, an operation that may degrade the connection performance and lead to network~disconnections.

Motivated by these considerations, in this study we investigate the reliability and efficiency of  \gls{v2n} communication when using mmWave and LTE technologies. 
More specifically, we conducted an extensive simulation campaign to 
(i) compare the performance of the \gls{mmwave} and the \gls{lte} technologies in terms of achievable data rate, communication stability and signal outage probability under realistic dynamic scenarios, and
(ii) assess their benefits and potential shortcomings in relation with target V2N application requirements. 
In order to have realistic movement of the vehicles in an urban environment, we generate some mobility traces using \gls{sumo}~\cite{SUMO2012}, an open  road traffic simulator designed to handle and model the traffic of large road networks.
The results show that the support of high-frequency bands has the potential to guarantee extremely high throughput, and that LTE transmissions enable  stable  communications.
We conclude that the parallel use of  different \glspl{rt} makes it possible to complement the limitations of each type of network and therefore represents a viable approach to guarantee both  high-capacity and resilient \gls{v2n} communications.

The rest of the paper is organized as follows.
In Sec.~\ref{sec:related_work} we review some of the most relevant contributions related to \gls{v2n} communication, while
in Sec.~\ref{sec:rats_to_support_v2i_networking} we overview the strengths and weaknesses of the LTE and the mmWave paradigms as enabling technologies for vehicular communications.
 The system and mobility models and the simulation parameters are described in Sec.~\ref{sec:system_model_and_simulation_parameters}, while in Sec.~\ref{sec:comparative_results} we present our main findings and simulation results. 
 Finally, conclusions and suggestions for future work are provided in Sec.~\ref{sec:conclusion_and_future_work}.

\begin{table*}[!t]
\centering
\caption{Description of the characteristics of radio interfaces currently being considered for \gls{v2n} communications. }
\label{tab:rats}
\renewcommand{\arraystretch}{1}
\begin{tabular}{p{2cm}p{6.3cm}p{6.3cm}}
\toprule
RT & Pros & Cons \\
\midrule
LTE \cite{3GPP_LTE,araniti2013LTE} &
\begin{tabular}[t]{@{\textbullet~}p{5.8cm}@{}}
 Low latency (RTT within a cell lower than 10 ms)\\
 Wide deployment\\
 Omnidirectional/broadcast transmissions\\
 Scheduled transmissions
\end{tabular} &
\begin{tabular}[t]{@{\textbullet~}p{5.8cm}@{}}
 Scalability issues\\
 Non-ubiquitous coverage\\
 Limited transmission data rate	
\end{tabular} \\
\midrule
mmWaves \cite{rappaport2014millimeter} &
\begin{tabular}[t]{@{\textbullet~}p{5.8cm}@{}}
 Very large bandwidth ($>$50 times more than LTE)\\
 Beamforming gain\\
 Spatial isolation\\
 Inherent security and privacy
\end{tabular} &
\begin{tabular}[t]{@{\textbullet~}p{5.8cm}@{}}
 Very large path loss\\
 Signals do not penetrate through solid material\\
 Need to set up aligned transmissions\\
 Significant shadowing, reflection and scattering
\end{tabular} \\
\bottomrule
\end{tabular}
\end{table*}


\section{Related Work} 
\label{sec:related_work}
The application of   mmWave technology in the automotive context is not new and some preliminary studies have already demonstrated the feasibility of designing mmWave-based  protocols to support vehicular communications~\cite{kato2001its,yamamoto2008path}.  
The research has also focused on  the design of automotive radars operating in the 77 GHz band to enable driver assistance functionalities~\cite{hasch2012millimeter}.
More recently, the authors in~\cite{shah20185g,lu2014connected}   
have explored the benefits of emerging technologies (e.g., 5G communications, network slicing, mobile edge computing) to support advanced automotive~applications.

Despite this growing interest, the severe radio conditions resulting from the mobility of vehicles and their relative speed with respect to the road infrastructures, and the need to maintain beam alignment between the communication endpoints, pose significant challenges to the design of \gls{v2n} systems operating at \glspl{mmwave} ~\cite{MOCAST_2017}. A stochastic analysis of the connectivity probability in mmWave networks is presented in \cite{giordani2018coverage}, where the authors derive a probabilistic connectivity model that makes it possible to study the communication performance when varying the nodes speed, the beam alignment periodicity, the infrastructure density and the antenna geometry, under some simplifying assumptions. The problem of beam alignment is addressed in some other recent studies, e.g.,~\cite{beamDesignV2I_Heath}, which proposes sophisticated strategies to reduce the beam alignment overhead under dynamic networks and unpredictable wireless channel evolution, and \cite{locationAided_Garcia}, where a location-aided hierarchical beamforming approach is proposed to achieve ultra-fast alignment and connection. 

However, given the variety of the automotive services and the heterogeneity of their requirements, it is unlikely that \gls{v2n} communications can be supported by a single radio access interface, rather the orchestration of multiple \glspl{rt}, which complement each other's capabilities, might be the solution. 
Although the effectiveness of such mixed solutions has already been demonstrated (e.g., \cite{sepulcre2015context,dressler2014inter}), at the time of writing  there  is no prior work that quantitatively investigates the strengths and weaknesses of the enabling technologies and, in particular, of \gls{mmwave} and  \gls{lte}, which are currently considered for future \gls{v2n} communications. This paper contributes to partially fill this gap. 


\section{Radio Technologies to Support \\ V2N Communications} 
\label{sec:rats_to_support_v2i_networking}
The 3GPP has provided a basic set of  V2N connection requirements to support classic services, mainly intended to avoid or mitigate vehicle accidents, but also to enable a wide range of other safety,  environmental and societal benefits~\cite{3GPP_22886}. 
%
%
%
%
These safety-related services have very stringent demands in terms of communication reliability and stability, while relatively low data rates are usually~expected. 

However, \gls{v2n} communications will also enable more advanced services, e.g., by sharing sensory data to build collective context awareness and permitting the vehicles to enhance their perception beyond what provided by the onboard instrumentation. The digital horizon built by vehicles can be exploited to coordinate their maneuvers and support safe and automated driving applications,  which are collectively referred to as \emph{cooperative driving/maneuvering}. 

Finally, \gls{v2n} transmissions enable \emph{infotainment} applications, which generically refer to a set of services that deliver a combination of information and entertainment. 
For media download applications (e.g., Internet browsing) the required throughput will depend on the content type, while latency is reasonably tolerated because the download can be generally completed within some flexible time frame. Video streaming, on the other hand, demands stable throughput (especially for high-quality video) with stringent latency. Other real-time services (e.g., on-line gaming) generate data streams with relatively low bitrate but require the dynamic maintenance of  multicast communication, reliable connections and minimum end-to-end delays.

Based on the above considerations, in this section  we overview the characteristics of candidate RTs currently being considered to support
\gls{v2n} communications, i.e.,  the LTE standard and the mmWave technology,
and discuss their possible shortcomings in relation with the aforementioned application requirements. Table~\ref{tab:rats} provides a short summary of this discussion.

\subsection{LTE Communications} 
\label{sub:lte_communications}

The \gls{lte} standard \cite{3GPP_LTE} operates in the sub-6 GHz spectrum.
 It can provide a round-trip-time within a cell theoretically
lower than 10 ms.
Moreover, the \gls{lte} infrastructure is widely deployed,  the transmissions are omnidirectional and \gls{lte} has the potential to support multicast and
broadcast data distribution~\cite{araniti2013LTE}.
Finally, \gls{lte} transmissions are scheduled, so that collisions are avoided and mutual interference is minimized.

However, there are several challenges that need to be addressed
before LTE can be massively exploited in vehicular environments.
First, access and transmission latencies increase with the number of vehicles in the cell, thus raising scalability issues.
Second, despite the almost ubiquitous coverage of the \gls{lte} infrastructures, still the connection may not be always available (e.g., in tunnels or in rural areas).
Third, \gls{lte} offers limited downlink capacity (i.e., around 300 Mbps in Release 8, though much lower rates are typical in high mobility scenarios), which may not be sufficient to support  some categories of  applications (e.g., advanced driving or infotainment). Finally, the \gls{v2n} communications should share the LTE cell capacity  with classic human-initiated services, which may result in a strong performance degradation in either (or both) traffic~categories. 

In conclusion, the \gls{lte} standard, despite its clear strengths, may not be able to support all future \gls{v2n} requirements.

\subsection{Millimeter Wave Communications} 
\label{sub:millimeter_wave_communications}

The mmWave spectrum between 10 GHz and 300 GHz will be a cornerstone of next-generation wireless networks (5G)~\cite{RanRapE:14} and represents a promising candidate to support the demands  of future automotive applications.
As mentioned, in fact, the large bandwidth available at mmWave frequencies permits a much higher bitrate than  traditional systems. Moreover, the small wavelengths make it practical to design very large antenna arrays to obtain high beamforming gains~and spatial isolation that, in turn, decrease the interference among different vehicles and have the potential to reduce the risk of eavesdropping~\cite{rappaport2014millimeter}\footnote{Although security in automotive systems will mainly rely on bit-level cryptographic techniques, mmWave transmissions leveraging \emph{physical layer security} look promising to mitigate computational and communication overhead for application-level encryption~\cite{wang2016physical}.}.

On the other hand, there are still a number of challenges that need to be addressed before enabling \gls{mmwave} V2N solutions. To begin with, mmWave signals do not penetrate most solid materials, so that shadowing and blockage phenomena may be quite frequent. Moreover, the path loss at \glspl{mmwave} is very large and, hence, long-range omnidirectional radio  transmissions are not possible, which makes data broadcasting more complex. 
Such a severe path loss is counteracted by adopting directional communication that, however, requires precise beam alignment between transmitter and receiver, whose maintenance may imply control overhead and latency. 
These challenges are further exacerbated in highly dynamic vehicular scenarios, as the beam alignment may be lost before a data exchange is completed~\cite{giordani2018coverage}. 
Although the alignment overhead can be reduced by the dissemination of context information, e.g., geographical position of vehicles and \glspl{rsu}, or by implementing digital beamforming architectures,  the direct applicability of the \gls{mmwave} technology to a vehicular context is still not clear and has become a research focus in the area of intelligent autonomous systems.


%

\section{System Model and Simulation Parameters} 
\label{sec:system_model_and_simulation_parameters}

In this section we present the system model we considered to evaluate the  performance of the \gls{v2n} system.
The channel and mobility models are described in Secs.~\ref{sub:channel_model} and~\ref{sub:mobility_model}, respectively, while the
 simulation~parameters are illustrated in Sec.~\ref{sub:simulation_parameters}.

\subsection{Channel Model} 
\label{sub:channel_model}
In \gls{v2n} systems, we reasonably expect that  other vehicles, pedestrians or urban buildings can
block the link connecting the target vehicle and its serving \gls{rsu}.
It is therefore necessary to distinguish between
\gls{los} and \gls{nlos} nodes.\\

\textbf{LTE Model.}
For the LTE cells, we consider  the  3GPP specifications~\cite{3GPP_LTE_channel} for an outdoor dense scenario. 
Accordingly, a vehicle at distance $d$ from its LTE serving infrastructure will be in \gls{los} with probability
\medmuskip=0mu
\thickmuskip=0mu
\eqPLLTE
\medmuskip=6mu
\thickmuskip=6mu
and in \gls{nlos} with complementary probability $P_{\rm NLOS}(d) = 1-P_{\rm LOS}(d)$.
The path loss for the two conditions, in dB, is given by
 \begin{empheq}[left = \empheqlbrace]{align}
&PL_{\rm LOS}(d) = 103.4+24.2\log_{10}(d)\;;\\
&PL_{\rm NLOS}(d) = 131.1+42.8\log_{10}(d)\;.
\end{empheq}
In addition, we consider a fast Rayleigh fading, which is modeled as a stochastic gain with unit power (in linear scale). 
 	
\textbf{Millimeter Wave Model.}
As already mentioned, the  short wavelengths of
\gls{mmwave} signals  result in significant
shadowing, reflection and diffusion~\cite{geng2009millimeter}, especially  where the coverage frequently implies \gls{nlos} links.
We therefore adopt the channel model described in ~\cite{Mustafa} that provides a realistic assessment of the \gls{mmwave} network in a dense urban~deployment. 
According to this model, 
the instantaneous probability to be in \gls{los} and \gls{nlos} at distance $d$ from the transmitter is equal to $P_{\text{LOS}}(d) = e^{-a_{\text{LOS}}d}$ and $P_{\text{NLOS}}(d) = 1-P_{\text{LOS}}(d)$, respectively, where   ${a_{\text{LOS}} = 0.0149}$~m$^{-1}$.
 The path loss is computed~as:
 \begin{empheq}[left = \empheqlbrace]{align}
&PL_{\rm LOS}(d) = 61.4 + 2\log_{10}(d) + \xi_{\sigma_L} \;; \\
&PL_{\rm NLOS}(d) = 72 + 2.92\log_{10}(d) + \xi_{\sigma_N}\;;
\end{empheq}
where  ${\xi_{\sigma_L}\sim N(0,\sigma_L^2)}$ and ${\xi_{\sigma_N}\sim N(0,\sigma_N^2)}$ characterize the shadowing (in dB) in \gls{los} and \gls{nlos} conditions, respectively, with $\sigma_L = 5.8$ dB and $\sigma_N=8.7$ dB at 28~GHz~\cite{Mustafa}.

\subsection{Mobility Model} 
\label{sub:mobility_model}

\begin{figure}[t!]
	\centering
		\includegraphics[width=0.8\columnwidth]{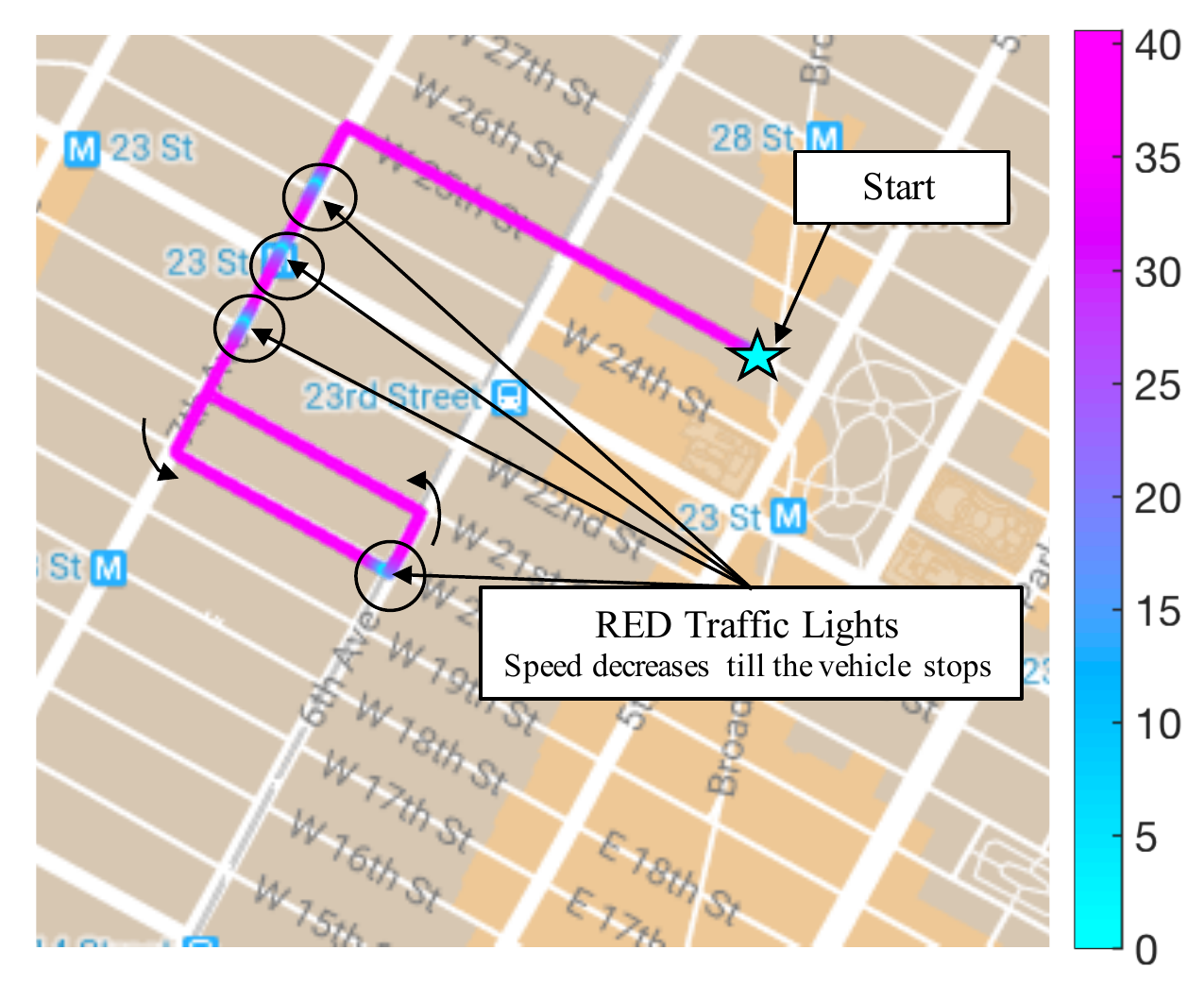}
		\caption{The car's instantaneous speed trace (km/h) overlaid to the considered  NYC road map. The car stops four times between  24th, 23rd, 22nd streets and 7th avenue and between 20th street and 6th avenue due to red traffic lights.}
		\label{fig:sumo_scenario}
\end{figure}

For our simulations, we use a real road map data imported from \textit{Openstreetmaps} (OSM) \cite{websiteopenstreetmap}, an open-source tool which
combines wiki-like user generated data with
free access information, allowing users to create editable maps of the world.
We consider the OSM map of New York City and, more specifically, of the Madison Square Park district between  W~16th and W~27th streets.

Moreover, in order to consider realistic mobility models and representative speed traces, 
we  simulated the mobility of cars using  \gls{sumo}~\cite{SUMO2012}, a
powerful, open-source traffic generator that supports the modeling of intermodal traffic systems including road vehicles and structures, public transports and pedestrians.
The vehicles move through the street network (imported from OSM) according to a \texttt{randomTrip} mobility model, which generates
trips with random origins and destinations, and speed values which depend on the interaction of the moving vehicle with the road and network elements (e.g., traffic lights or road intersections), as displayed in Fig.~\ref{fig:sumo_scenario}.

\subsection{Simulation Parameters} 
\label{sub:simulation_parameters}
The simulation parameters are based on realistic system design considerations and are summarized in Table~\ref{tab:params}.
The mmWave and LTE \glspl{rsu} are deployed over an area of 1 km$^2$ (which is deemed sufficient to avoid undesired boundary effects) according to two independent   Poisson Point Processes (PPPs) with densities $\lambda_{\rm mmW}$ and $\lambda_{\rm LTE}$, respectively.
In particular, according to the 3GPP specifications~\cite{3GPP_LTE_channel}, the LTE macro inter-site distance is 500 m, i.e., $\lambda_{\rm LTE} = 4$ RSU/km$^2$, while $\lambda_{\rm mmW}$ varies from 10 to 100  RSU/km$^2$ (the trade-off oscillates between better coverage and increased deployment cost).
We assume the LTE systems operate in the  2 GHz legacy band, with 20~MHz of bandwidth and omnidirectional~transmission. 
    \begin{figure*}[b!]
     \centering
     \renewcommand\thefigure{3}
     		\setlength{\belowcaptionskip}{-0.3cm}
	\setlength\fwidth{0.9\textwidth}
	\setlength\fheight{0.19\textwidth}
	\includegraphics[width=0.99\textwidth]{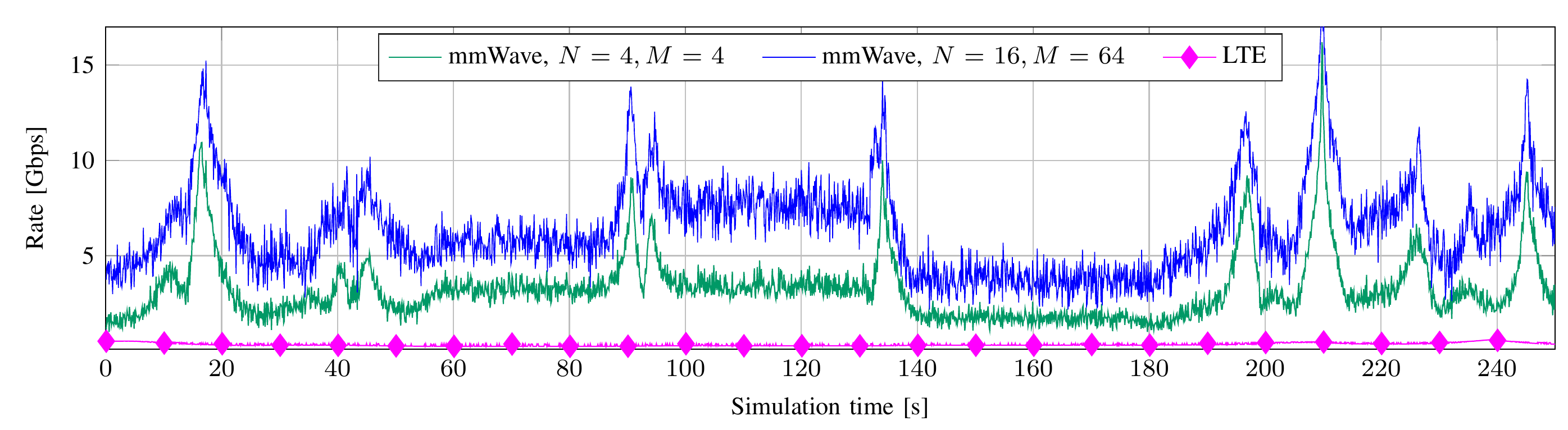}
    \caption{\footnotesize Example of time-varying rate for LTE (diamond markers) and \gls{mmwave} communications (solid plain lines), with different antenna array configurations, perfect beam alignment, and $\lambda_{\rm mmW}=100$ RSU/km$^2$.}
    \label{fig:rate_beam}
      \end{figure*} 
For the mmWave links, instead, we assume to operate at 28~GHz, with 1~GHz bandwidth and (in general) directional transmission. To this end, vehicles and \glspl{rsu} are equipped with Uniform Planar Arrays (UPAs) of $N$ and $M$ elements, respectively, allowing to steer beams consisting of a main lobe of predefined width (which depends on $N$ and $M$) and a side lobe that covers the remainder of the antenna radiation pattern.\footnote{There exists a strong correlation among number of antenna elements, beam width, and beamforming gain: the more antenna elements in the system, the narrower the beams, the more  directional the transmission, and the higher the gain that can be achieved by beamforming.} For the sake of completeness,  we also consider omnidirectional \gls{mmwave} transmissions at the vehicle side, which corresponds to $N=1$. Notice that, in general, $M\geq N$, since \glspl{rsu} are usually less space-constrained than~cars. 

\begin{figure}[t!]
\renewcommand\thefigure{2}
     \centering
     		\setlength{\belowcaptionskip}{-0.3cm}
	\setlength\fwidth{0.62\columnwidth}
	\setlength\fheight{0.55\columnwidth}
%
%
\definecolor{mycolor1}{rgb}{1.00000,0.00000,1.00000}%
\pgfplotsset{
tick label style={font=\footnotesize},
label style={font=\footnotesize},
legend  style={font=\footnotesize}
}
\begin{tikzpicture}

\begin{axis}[%
width=\fwidth,
height=\fheight,
at={(0\fwidth,0\fheight)},
scale only axis,
xmin=10,
xmax=100,
xlabel style={font=\color{white!15!black}},
xlabel={$\lambda_{\rm mmW}$ [RSU/km$^2$]},
ymin=0,
ymax=9,
ylabel style={font=\color{white!15!black}},
ylabel={Rate [Gbps]},
axis background/.style={fill=white},
label style={font=\footnotesize},
xmajorgrids,
ymajorgrids,
yminorgrids,
legend columns={2},
legend style={at={(1.2,1.15)},legend cell align=left, align=center, draw=white!15!black},
]
\addplot [color=blue, line width=0.5pt, mark size=2.3pt, mark=o, mark options={solid, blue}]
  table[row sep=crcr]{%
10	6.60838050947632e-12\\
20	0.002890271288569\\
30	1.16720499740272\\
40	3.45873676399393\\
50	4.63725452733742\\
60	5.12028723710665\\
70	5.35646306386931\\
80	5.39041242511946\\
90	5.41292122090815\\
100	5.33270863988879\\
};
\addlegendentry{mmWave, $N = 1, M=64$}

\addplot [color=blue, line width=0.5pt, mark size=2.3pt, mark=square, mark options={solid, blue}]
  table[row sep=crcr]{%
10	3.42126327268488e-12\\
20	0.000979694417163057\\
30	0.516246098388879\\
40	2.38068831876754\\
50	3.49296406091064\\
60	4.09469803921203\\
70	4.27326862650114\\
80	4.33336538695909\\
90	4.34118239865935\\
100	4.22179779721566\\
};
\addlegendentry{mmWave, $N = 4, M=4$}

\addplot [color=blue, line width=0.5pt, mark size=2.3pt, mark=triangle, mark options={solid, blue}]
  table[row sep=crcr]{%
10	1.96351823689156e-10\\
20	0.0230201153744074\\
30	4.02370551292305\\
40	6.76908849714244\\
50	7.91284554676202\\
60	8.2103751460876\\
70	8.19451143488221\\
80	8.21012904178305\\
90	8.06961907096686\\
100	8.0016700363652\\
};
\addlegendentry{mmWave, $N = 16, M=64$}

\addplot [color=mycolor1, mark size=3.5pt, mark=diamond*, mark options={solid, mycolor1}]
  table[row sep=crcr]{%
10	0.323244809377515\\
20	0.322413264818976\\
30	0.324080124218444\\
40	0.324441036598499\\
50	0.324214262328454\\
60	0.324041843059434\\
70	0.323585650526375\\
80	0.324919350058648\\
90	0.324434155571032\\
100	0.324994522734308\\
};
\addlegendentry{LTE}

\end{axis}
\end{tikzpicture}%
    \caption{Average achievable data rate when varying $\lambda_{\rm mmW}$ for \gls{mmwave} communications with perfect tracking and different antenna array configurations (empty markers).  A fixed LTE configuration (full diamond markers) is considered.}
    \label{fig:datarate}
      \end{figure}

For the beam alignment, we consider the procedure described in~\cite{giordani2018coverage}, according to which the nodes exchange channel quality reports at the beginning of every slot of duration $T_{\rm tr}$, so that vehicles and \glspl{rsu} can regularly identify the optimal directions for their respective beams. 
In order to minimize the tracking overhead (which might be significant when considering very low values of $T_{\rm tr}$), in  case the alignment is lost within a slot,  connectivity can  be recovered only at the beginning of the subsequent slot, i.e., when a new beam alignment operation will be completed. In order to get insights on some limit performance, we will also consider the \emph{perfect beam alignment} case where the beam alignment procedure is performed continuously ($T_{\rm tr}=0$) and the beams are hence always perfectly aligned (e.g., thanks to the use of digital beamforming or other external means).
\renewcommand{\arraystretch}{0.9}
\begin{table}[!t]
\small
\centering
\caption{Main simulation parameters.}
\begin{tabularx}{0.98\columnwidth}{ @{\extracolsep{\fill}} lll}
\toprule
Parameter & Value & Description \\ \midrule
$P_{\rm TX,mmW}$ & $30$ dBm & mmWave TX power    \\
$W_{\rm mmW}$ & $1$ GHz & mmWave bandwidth \\
$f_{\rm c, mmW}$ & $28$ GHz &  mmWave carrier frequency \\
$\lambda_{\rm mmW}$ & $\{10,..,100\}$ RSU/km$^2$  & mmWave RSU density\\
$T_{\rm tr}$ & $\{0.1,1\}$ s & Tracking periodicity\\
$N$ & $\{1,4,16\}$  & Vehicle  array size  \\
$M$ & $\{4,16,64\}$  & mmWave RSU array size  \\
\midrule
$P_{\rm TX,LTE}$ & $46$ dBm & LTE TX power    \\
 $W_{\rm LTE}$ & $20$ MHz & LTE bandwidth \\
$f_{\rm c, DSRC}$ & $2$ GHz &  LTE carrier frequency \\
$\lambda_{\rm LTE}$ & $4$ RSU/km$^2$  & LTE RSU density\\
\midrule
\multicolumn{3}{c}{LTE channel parameters ~$\sim$\cite{3GPP_LTE_channel} }  \\
\multicolumn{3}{c}{mmWave channel parameters ~$\sim$\cite{Mustafa} }  \\
\bottomrule
\end{tabularx}
\label{tab:params}
\end{table}       

The statistical results are derived through a Monte Carlo approach,
where multiple independent simulations are repeated to get
different statistical quantities of interest.
More specifically, we
analyze the received signal strength between a single target transmitting vehicle
and its serving infrastructure for different values of~$\lambda_{\rm mmW}$. 
The signal strength is also related to the achievable data rate, which is calculated according to the Shannon formula.
Given a bandwidth $W$ (either $W_{\rm mmW}$ or $W_{\rm LTE}$, according to the technology being considered),  and the SNR $\Gamma_{i,j}$ between the transmitter $i$ and its candidate receiver $j$, the average (Shannon) data rate is therefore computed as 
\begin{equation}
	R_{\rm Sh} = W \log_2(1 + \Gamma_{i,j}).
	\label{eq:datarate}
\end{equation}

\section{Comparative Results} 
\label{sec:comparative_results}
In the following subsections we provide some numerical results to compare the connectivity performance of LTE and mmWave-based \gls{v2n} transmissions, which will be evaluated in terms of achievable data rate, stability and outage probability.

In Fig.~\ref{fig:datarate} we compare the two technologies in terms of average data rate experienced by the target vehicle assuming perfect beam alignment for the mmWave links and different antenna configurations (including omnidirectional transmissions at the vehicular node, $N=1$). The data rate is computed according to Eq.~\eqref{eq:datarate}.
We observe that the very large bandwidth available to the mmWave systems (50 times larger than in LTE) provides much higher throughput, which may be required by some \gls{v2n} applications (e.g., high-resolution video streaming).
Moreover, even with an omnidirectional mmWave antenna at the vehicular node, the connection can still provide acceptable average bitrate, provided that the mmWave \glspl{rsu} are sufficiently dense (to increase the \gls{los} probability). Finally, we note that by further increasing $\lambda_{\rm mmW}$, the mutual interference among the \glspl{rsu} will eventually impact the average bitrate, which will start decreasing.

In Fig.~\ref{fig:rate_beam} we plot the evolution of the transmit rate experienced by a moving vehicle in a specific \gls{sumo} simulation of duration $T_{\rm sim}= 250$~s, for two antenna configurations.
It is apparent that the mmWave throughput  depends on the array factors, i.e., on the beamforming gain. 
From Fig~\ref{fig:rate_beam}, it is also interesting to observe that the rate traces follow accurately the movements of the car along the streets. For example, the almost constant rate  intervals (e.g., from  $60$ s to $90$ s, from $100$ s to $130$ s, and from $140$ s to $190$ s) correspond to stopping periods at the red traffic lights, while the  peaks correspond to the crossing of the coverage range of the mmWave \glspl{rsu}, whose range is geometrically determined by their beamwidths.
Finally, we observe that the mmWave rate presents clear fluctuations and variations, which are mainly due to transitions from LOS to NLOS, and \emph{vice versa}, and to the small scale fading.

\begin{figure}[t!]
     \centering
     		\setlength{\belowcaptionskip}{-0.5cm}
      		\setlength{\belowcaptionskip}{0cm}
	\setlength{\belowcaptionskip}{0cm}
	\setlength\fwidth{0.62\columnwidth}
	\setlength\fheight{0.55\columnwidth}
	\includegraphics[width=0.99\columnwidth]{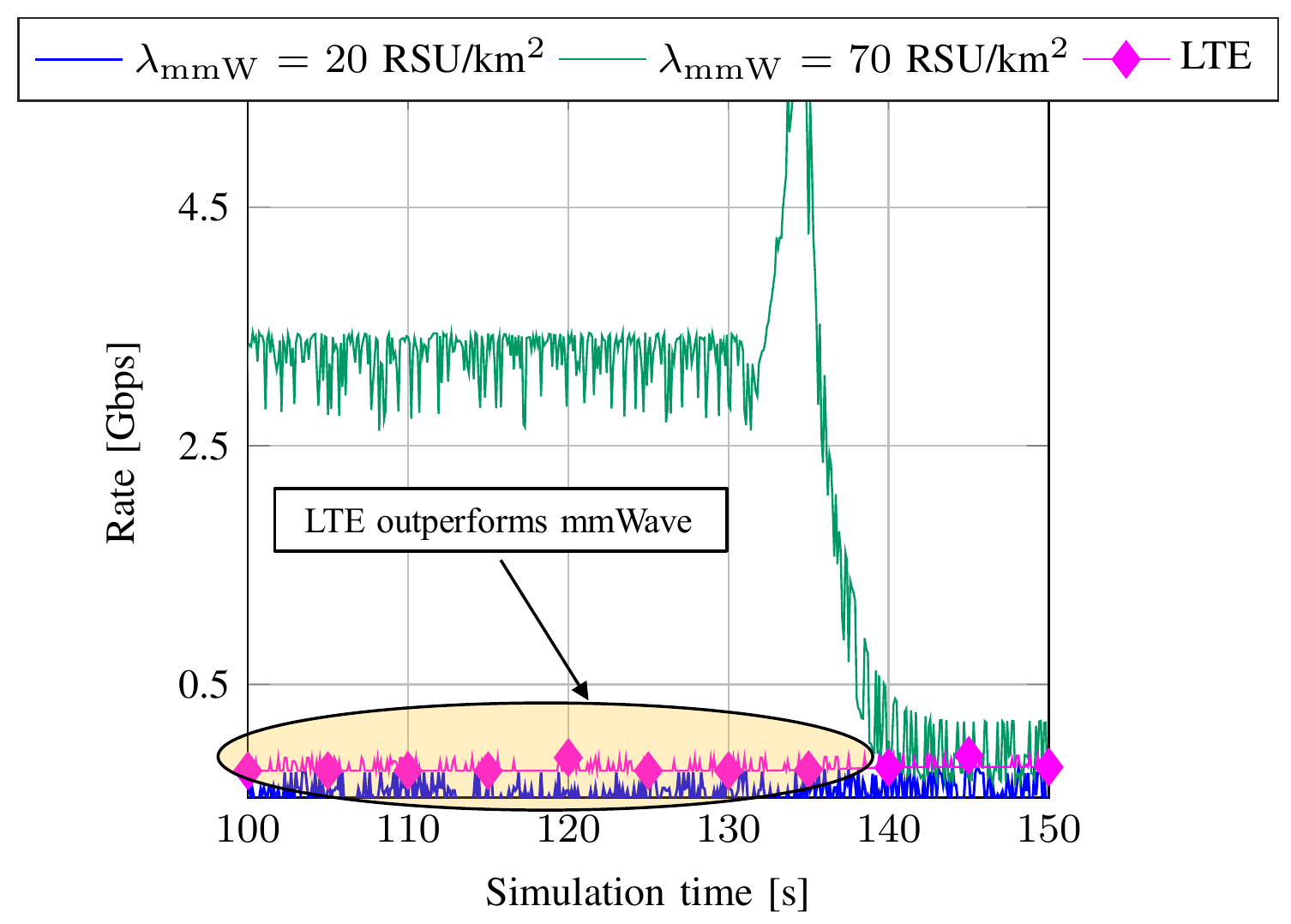}
    \caption{\footnotesize  Example of LTE and mmWave rate over time, for different values of the \gls{rsu} density ($\lambda_{\rm mmW}=20$ and $70$ RSU/km$^2$), perfect beam alignment, and $N=4$ and~$M=4$.}
    \label{fig:rate_density}
      \end{figure}

In Fig.~\ref{fig:rate_density} we can observe the effect  on the transmission rate of different mmWave \gls{rsu} deployment strategies. 
As expected, the rate decreases in sparser networks (i.e., with $\lambda_{\rm mmW}=20$ RSU/km$^2$) since vehicles generally need to connect with farther \glspl{rsu}. 
In these conditions,  LTE can provide better performance (despite the low density of LTE \glspl{rsu}, equal to $\lambda_{\rm LTE}=4$ RSU/km$^2$).

\subsection{Stability Performance} 
\label{sub:robustness_performance}

\begin{figure}[t!]
     \centering
     		\setlength{\belowcaptionskip}{-0.3cm}
	\setlength\fwidth{0.63\columnwidth}
	\setlength\fheight{0.55\columnwidth}
%
%
\definecolor{mycolor1}{rgb}{0.20810,0.16630,0.52920}%
\definecolor{mycolor1}{rgb}{1.00000,0.00000,1.00000}%
\pgfplotsset{
tick label style={font=\footnotesize},
label style={font=\footnotesize},
legend  style={font=\footnotesize}
}

\begin{tikzpicture}

\begin{axis}[%
width=\fwidth,
height=\fheight,
at={(0\fwidth,0\fheight)},
scale only axis,
bar shift auto,
xmin=0,
xmax=6,
xtick={1,2,3,4,5},
xticklabels={10,30,50,70,90},
xlabel style={font=\color{white!15!black}},
xlabel={$\lambda_{\rm mmW}$ $ \text{ [RSU/km}^2]$},
ylabel style={font=\color{white!15!black}},
xlabel style={font=\color{white!15!black}},
ymin=0,
ymax=1.3,
ylabel={$\rho_{\rm var}$},
label style={font=\footnotesize},
axis background/.style={fill=white},
xmajorgrids,
ymajorgrids,
legend style={legend cell align=left, align=left, at={(1,0.95)},  draw=white!15!black, /tikz/every even column/.append style={column sep=0.35cm}, legend columns = 1}]

 1.2341    0.8447    0.6760    0.5973    0.5492    0.5209    0.5106    0.4903    0.4900    0.4900

\addplot[ybar, bar width=0.229, fill=blue, draw=black, area legend] table[row sep=crcr] {%
1	1.2341\\
2	0.6760  \\
3	0.5492	\\
4	0.5106	\\
5	0.4900	\\
};
\addlegendentry{mmWave, $N=16,M=64$}

\addplot [color=mycolor1, mark size=3.5pt, mark=diamond*, mark options={solid, mycolor1}]
  table[row sep=crcr]{%
0	0.2484\\
1	0.2484\\
2	0.2484\\
3	0.2484\\
4	0.2484\\
5	0.2484\\
6	0.2484\\
};
\addlegendentry{LTE}

\end{axis}
\end{tikzpicture}%
    \caption{\footnotesize LTE and \gls{mmwave} stability index $\rho_{\rm var}$ when varying $\lambda_{\rm mmW}$, with perfect beam alignment and $N=16$ and $M=64$. }
    \label{fig:variability}
      \end{figure}
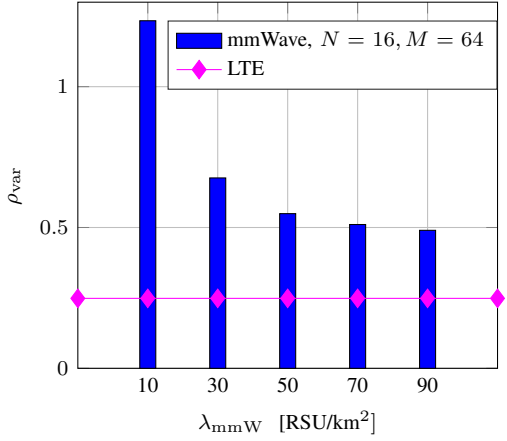

Following the analysis we proposed in~\cite{JSAC_2017}, we measure the \emph{stability} performance of a link in terms of the coefficient of variation of the transmission rate, which is defined as 
\begin{equation}
\rho_{\rm var} = \frac{\mathrm{STD}(R)}{\mathbb{E}[R]},	
\label{eq:var}
\end{equation}
where $\mathrm{STD}(R)$ and $\mathbb{E}[R]$ are the standard deviation and mean, respectively, of the throughput $R$ experienced by a certain \gls{rt}. 
High values of $\rho_{\rm var}$ indicate significant channel instability, meaning that the rate  would be affected by local variations and periodic degradations.
Let $\rho^{\rm mmW}_{\rm var}$ and $\rho^{\rm LTE}_{\rm var}$ denote the stability indices \eqref{eq:var} for mmWave and LTE links, respectively.
From Fig.~\ref{fig:variability} we observe that $\rho^{\rm mmW}_{\rm var} > \rho^{\rm LTE}_{\rm var}$ for all values of $\lambda_{\rm mmW}$, which reflects the much higher variability of the performance guaranteed by mmWave links as a result of the scattering from nearby buildings, vehicles and terrain surfaces. These variations may actually hinder the adoption of mmWave links to support some types of \gls{v2n} applications, especially those requiring long-term stable throughput (e.g., real-time services).
However, the stability of the mmWave links increases in dense environments, as the probability of path loss outage decreases.

\begin{figure}[t!]
     \centering
     		\setlength{\belowcaptionskip}{-0.3cm}
	\setlength\fwidth{0.63\columnwidth}
	\setlength\fheight{0.55\columnwidth}
	\input{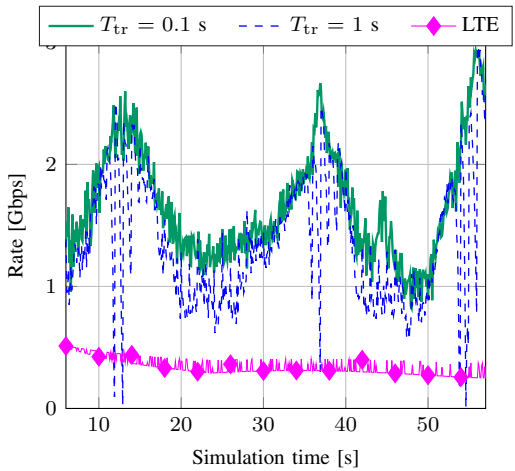}
    \caption{\footnotesize  Example of time-varying LTE and \gls{mmwave} rates for different values of the tracking period ($T_{\rm tr}=0.1$~s and $T_{\rm tr}=1$~s), and with $\lambda_{\rm mmW}=30$ RSU/km$^2$, $N=16$ and $M=64$.}
    \label{fig:rate_Ttr}
      \end{figure}

In case of directional mmWave transmissions, the misalignment between the transmitter and the receiver can also significantly impact the link stability, as exemplified in Fig.~\ref{fig:rate_Ttr} which shows the rate achieved by the mmWave link when changing the beam tracking period. As $T_{\rm tr}$ increases, the rate presents significant variations which, in some cases, may lead to network disconnections. 
Conversely, the omnidirectional transmissions of LTE systems offer more stable connectivity and, under some circumstances, even higher throughput than mmWave strategies.

\begin{figure}[t!]
     \centering
      		\setlength{\belowcaptionskip}{-0.3cm}
	\setlength\fwidth{0.63\columnwidth}
	\setlength\fheight{0.55\columnwidth}
	\includegraphics[width=0.75\columnwidth]{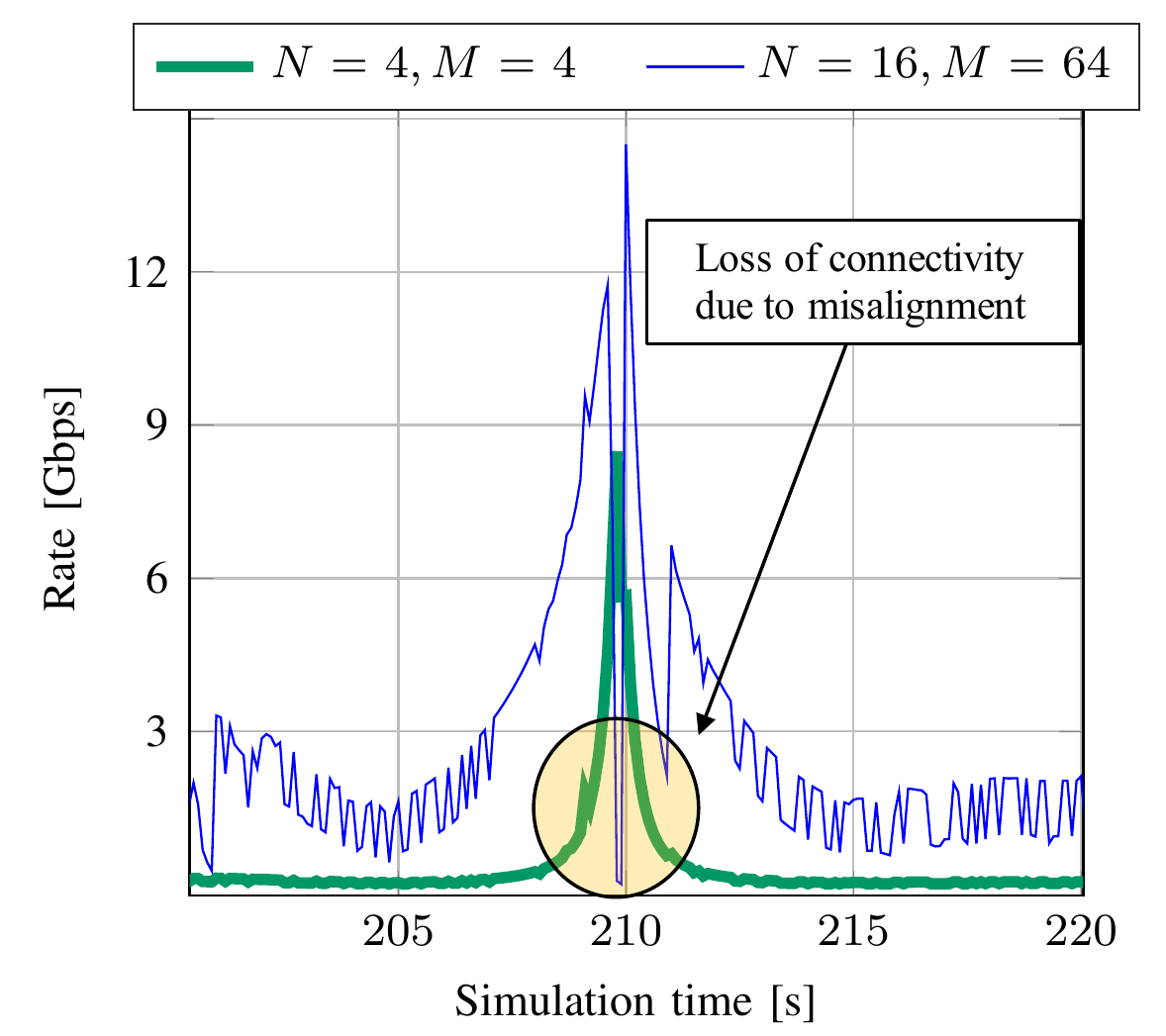}
    \caption{\footnotesize  Connection stability for different antenna array configurations considering mmWave propagation, with tracking period $T_{\rm tr}=1$~s and $\lambda_{\rm mmW}=100$ RSU/km$^2$.}
    \label{fig:rate_robustness}
      \end{figure}

The extent to which beam misalignment impacts the communication performance depends on several factors, including the beamwidth, as illustrated in Fig.~\ref{fig:rate_robustness}.
Wider beams provide more durable connections (even with very sporadic beam tracking), as they enlarge the area in which the vehicles can benefit from the coverage of their serving cells before disconnecting. 
In this context, however, the well-known stability versus data rate trade-off  arises: the narrower the beams, the higher but less stable the transmission rates.

\subsection{Outage probability} 
\label{sub:detection_accuracy_performance}

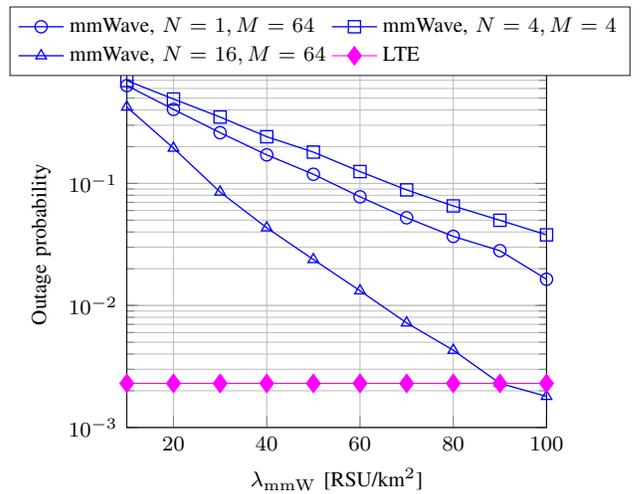
\begin{figure}[t!]
     \centering
     		\setlength{\belowcaptionskip}{-0.5cm}
      		\setlength{\belowcaptionskip}{0cm}
	\setlength{\belowcaptionskip}{0cm}
	\setlength\fwidth{0.63\columnwidth}
	\setlength\fheight{0.55\columnwidth}
%
%
\definecolor{mycolor1}{rgb}{1.00000,0.00000,1.00000}%
\pgfplotsset{
tick label style={font=\footnotesize},
label style={font=\footnotesize},
legend  style={font=\footnotesize}
}
\begin{tikzpicture}

\begin{axis}[%
width=\fwidth,
height=\fheight,
at={(0\fwidth,0\fheight)},
scale only axis,
xmin=10,
xmax=100,
ymode=log,
ymin=0.001,
ymax=1,
xlabel={$\lambda_{\rm mmW}$ [RSU/km$^2$]},
yminorticks=true,
axis background/.style={fill=white},
ylabel={Outage probability},
label style={font=\footnotesize},
xmajorgrids,
ymajorgrids,
yminorgrids,
legend columns={2},
legend style={at={(1.2,1.15)},legend cell align=left, align=center, draw=white!15!black},
]
\addplot [color=blue, line width=0.5pt, mark size=2.3pt, mark=o, mark options={solid, blue}]
  table[row sep=crcr]{%
10	0.6332\\
20	0.4044\\
30	0.2605\\
40	0.1715\\
50	0.1189\\
60	0.0777\\
70	0.0523\\
80	0.0368\\
90	0.0281\\
100	0.0164\\
};
\addlegendentry{mmWave, $N = 1, M=64$}

\addplot [color=blue, line width=0.5pt, mark size=2.3pt, mark=square, mark options={solid, blue}]
  table[row sep=crcr]{%
10	0.6957\\
20	0.4916\\
30	0.3499\\
40	0.2414\\
50	0.1806\\
60	0.1251\\
70	0.0885\\
80	0.0654\\
90	0.0499\\
100	0.0379\\
};
\addlegendentry{mmWave, $N = 4, M=4$}

\addplot [color=blue, line width=0.5pt, mark size=2.3pt, mark=triangle, mark options={solid, blue}]
  table[row sep=crcr]{%
10	0.4226\\
20	0.1947\\
30	0.085\\
40	0.0433\\
50	0.0238\\
60	0.0132\\
70	0.0072\\
80	0.0043\\
90	0.0023\\
100	0.0018\\
};
\addlegendentry{mmWave, $N = 16, M=64$}

\addplot [color=mycolor1, mark size=3.5pt, mark=diamond*, mark options={solid, mycolor1}]
  table[row sep=crcr]{%
10	0.0023\\
20	0.0023\\
30	0.0023\\
40	0.0023\\
50	0.0023\\
60	0.0023\\
70	0.0023\\
80	0.0023\\
90	0.0023\\
100	0.0023\\
};
\addlegendentry{LTE}

\end{axis}
\end{tikzpicture}%
    \caption{\footnotesize  Outage probability vs. $\lambda_{\rm mmW}$, considering both LTE and \gls{mmwave} communications. Perfect tracking is assumed. Different antenna array configurations are investigated.}
    \label{fig:outage}
      \end{figure}

 In Fig.~\ref{fig:outage} we evaluate the outage probability of the investigated V2N schemes, i.e., the probability that the received signal strength is below a predefined threshold.
We first observe that the outage probability for \gls{mmwave} transmissions decreases monotonically with the \gls{rsu} density $\lambda_{\rm mmW}$ (because of the closer distance between transmitter and receiver), and with the number of antenna elements at the nodes (because of the higher beamforming gain). 
Moreover, we notice that the outage probability of mmWave links degrades significantly when the vehicle has only one omnidirectional antenna ($N=1$), while LTE provides more reliable communications, with lower outage probability.

\section{Conclusion and Future Work} 
\label{sec:conclusion_and_future_work}

A limitation for the feasibility of  mmWave-enabled \gls{v2n} communication is the rapid dynamics of the mmWave channel and the high risk of misalignment between the communication endpoints caused by user mobility when using narrow beams. The vehicle may be suddenly in outage with respect to all the surrounding \glspl{rsu} and, in some scenarios, this may prevent the timely dissemination of data.
In this paper, we have compared the performance of LTE and  mmWave technologies in terms of achievable data rate, communication stability and outage probability. 
Our results showed that LTE systems provide limited data rates (i.e., up to a few hundreds of Mbps) but guarantee  stable and reliable transmissions thanks to the intrinsic stability of the low-frequency channels and the omnidirectional transmissions. 
Conversely, mmWave networks offer high-capacity but highly variable connectivity,
though their performance can be improved by considering very directional transmissions and frequent re-alignment operations, and by increasing the infrastructure density.

From our analysis we conclude that a practical way to complement the limitations of each type of network is via heterogeneous networking, i.e., by combining multiple radio access technologies into a single solution that is more robust and efficient than any individual approach.

The performance of  \gls{v2n} communications can be improved by designing advanced multi-\gls{rt} schemes able to instantaneously identify the best access solution to interconnect the vehicles and the infrastructure(s), or by envisioning synergistic combinations of the two approaches, e.g., exploiting LTE to help beam alignment of mmWave links, or using the two technologies in parallel to transmit layered content.   These research challenges are left for future work.

\bibliographystyle{IEEEtran}
\bibliography{bibliography.bib}

\end{document}